\def\spose#1{\hbox to 0pt{#1\hss}}
\def\approxlt{\mathrel{\spose{\lower 3pt\hbox{$\sim$}}
        \raise 2.0pt\hbox{$<$}}}
\def\approxgt{\mathrel{\spose{\lower 3pt\hbox{$\sim$}}
        \raise 2.0pt\hbox{$>$}}}

\def\multleft#1{\hbox to size{\vbox {\halign {\lft{##}\cr #1}}\hfill}\par}
\def\multright#1{\hbox to size{\vbox {\halign {\rt{##}\cr #1}}\hfill}\par}

\def\boxit#1{\vbox{\hrule\hbox{\vrule\kern3pt\vbox{\kern3pt
          #1 \kern3pt}\kern3pt\vrule}\hrule}}

\def\cm{{\rm\thinspace cm}}

\def\kpc{{\rm\thinspace kpc}}

\def\pc{{\rm\thinspace pc}}

\def\pcmsq{\hbox{$\cm^{-2}\,$}}

\documentclass[preprint]{aastex}

\received{}
\accepted{}

\slugcomment{Submitted to the Astrophysical Journal}

\shorttitle{\emph{Chandra} imaging of the Virgo cluster}
\shortauthors{Young, Wilson, \& Mundell}

\begin{document}

\title{\emph{Chandra} Imaging of the X-ray Core of the Virgo Cluster}

\author{A. J. Young, A. S. Wilson\altaffilmark{1}}

\affil{Astronomy Department, University of Maryland, College Park, MD
	20742}

\and

\author{C. G. Mundell\altaffilmark{2}}

\affil{Astrophysics Research Institute, Liverpool John Moores University, 12
 Quays House, Egerton Wharf, Birkenhead CH41 1LD, UK}


\altaffiltext{1}{Adjunct Astronomer, Space Telescope Science Institute, 3700
  San Martin Drive, Baltimore, MD 21218}

\altaffiltext{2}{Royal Society University Research Fellow}


\begin{abstract}

  We report sub-arcsecond X-ray imaging spectroscopy of M87 and the core of the
  Virgo cluster with the Chandra X-ray Observatory. The X-ray morphology shows
  structure on arcsecond ($\sim$ 100 pc) to ten  arcminute ($\sim$ 50 kpc)
  scales, the most prominent feature being an ``arc''  running from the east,
  across the central region of M87 and off to the southwest. A ridge in the
  radio map, ending in an ``ear''-shaped structure, follows the arc to the
  east. Depressions in the X-ray surface brightness correspond to the inner
  radio lobes and there is no evidence of shock-heated gas surrounding them. 
  There are also at least two approximately circular (centered near the
  nucleus) ``edges'' in the X-ray brightness distribution, the radii of which
  are slightly larger than the nuclear  distances of the inner radio lobes and
  intermediate radio ridges, respectively. We speculate that these
  discontinuities may be spherical pulses or ``fronts''  driven by the same jet
  activity  as is responsible for the radio structure; such pulses are found in
  recent numerical simulations.  All these results provide good evidence that
  the nuclear activity affects the intra-cluster medium. We present a
  temperature map of the intra-cluster medium, and obtain the temperature,
  pressure and cooling time as a function of nuclear distance for the arcs and
  the ambient intra-cluster medium. We show that the gas in the arcs is cooler
  than, and probably over-pressured with respect to, the ambient intra-cluster
  medium. The metal abundances of the cooler gas in the arc are somewhat
  enhanced relative to the ambient intra-cluster medium, favoring a ``buoyant
  plume'' origin for the X-ray arc, in which ambient gas near the nucleus is
  entrained by buoyant radio plasma and carried to larger nuclear distances.
  The gas within the inner ``front'' (nuclear distance $\simeq 3.5$~kpc $\simeq
  45\arcsec$) is at least a two-temperature plasma, with the cool component at
  $\simeq 1$~keV. This cool region is concentrated to the north of the nucleus
  and is strongly correlated with the H$\alpha$ + [NII] emission-line
  distribution.

\end{abstract}


\keywords{ galaxies: active --- galaxies: individual (M87) --- galaxies:
  cooling flows --- galaxies: clusters: individual (Virgo) --- galaxies:
  intergalactic medium --- X-rays: galaxies: clusters}


%

\section{Introduction}

The nearest X-ray emitting cluster of galaxies is the Virgo cluster in which
the X-ray emission is seen to be sharply peaked towards the cluster center
(Schreier, Gorenstein \& Feigelson 1982). In the central few tens of kpc the
temperature of the cluster gas decreases to $\approxlt 3$~keV (Canizares et al.
1979, 1982; Lea, Mushotzky \& Holt 1982; B\"ohringer et al. 1994; Nulsen \&
B\"ohringer 1995; B\"ohringer et al. 2001; Belsole et al. 2001) and within 10
kpc the cooling time is $\approxlt$ 1 Gyr (Stewart et al. 1984), suggesting the
presence of a 3 -- $20 M_\odot$ yr$^{-1}$ cooling flow (Lea et al. 1982;
Canizares et al. 1982). Simple cooling flow models have some problems, however,
such as the single temperature phase of the intra-cluster medium in Virgo
(B\"ohringer et al. 2001), and the observed absence of X-ray emitting gas with
temperature below 1~--~2~keV in other clusters (e.g. Schmidt, Allen \& Fabian
2001; Peterson et al. 2001; Fabian et al. 2001; Tamura et al. 2001). The giant
elliptical galaxy and powerful radio source M87 (Virgo A) lies at the core of
the Virgo cluster. At radio wavelengths M87 is seen to power spectacular jets
and lobes with significant structure extending over many tens of kpc (Owen,
Eilek \& Kassim 2000). Since M87 and the core of the Virgo cluster are nearby
(15.9 Mpc [Tonry 1991]), they have a large angular size on the sky ($1\arcsec =
77 \pc$) and present an ideal system in which to study the nature of
intra-cluster gas and its interaction with the radio source.

The interaction of the radio source with the cluster gas is thought to be
important because, at the present epoch, the inferred, model-dependent, kinetic
power in the jet is in the range $10^{43 - 44}$~erg~s$^{-1}$ (Bicknell \&
Begelman 1999), which may exceed the total X-ray luminosity of the cooling core
($10^{43}$ erg s$^{-1}$; Fabricant \& Gorenstein, 1983). Observations with
\emph{Einstein} (Feigelson et al 1987) revealed an ``arc'' of  X-ray emission
that is well correlated with the intermediate scale (nuclear distance $\simeq
15 \kpc \simeq 3 \arcmin$) radio ridges, and subsequent \emph{ROSAT}
(B\"ohringer et al. 1995; Harris, Biretta \& Junor 1999) and \emph{XMM-Newton}
(Belsole et al. 2001; B\"ohringer et al. 2001) observations find the gas in
this arc to be cooler than surrounding cluster gas. Since the X-ray and radio
features are quite well aligned, it is assumed there is some causal connection
between them. Models which may account for this connection include those of
Churazov et al. (2001), Br\"uggen \& Kaiser (2001) and Reynolds, Heinz \&
Begelman (2002, hereafter RHB), in which buoyant bubbles of radio plasma
interact with the intra-cluster medium. RHB also find that the transient
activity of the radio galaxy sends spherical ``pulses'' into the intra-cluster
medium.

We present here our first results on the cluster X-ray emission based on a 40
ks \emph{Chandra} observation of M87 and the core of the Virgo cluster. The
Galactic column density towards M87 is taken to be $N_H ({\rm Gal}) = 2.5
\times 10^{20} \pcmsq$ (Stark et al. 1992).

\section{X-ray Observations}

The core of the Virgo cluster was observed by \emph{Chandra} on 2000 June 29
for 40~ks with M87 at the aim-point of the Advanced CCD Imaging Spectrometer
(ACIS), on the S3 chip. After filtering for times of high background the total
usable exposure time was 36608~s. The ACIS background was obtained from a
compilation of relatively blank fields prepared by Maxim
Markevitch\footnote{\url{http://hea-www.harvard.edu/$^\sim$maxim/axaf/acisbg}},
and subtracted from our images and spectra. The images were corrected for
variations in the exposure time across the field caused by chip boundaries and
telescope vignetting. CIAO 2.1.3 and CALDB 2.7 were used to extract the data
and XSPEC 11.0.1 was used to model the spectra.

\section{Morphology} \label{sec:morphology}

At radio wavelengths M87 has a hierarchical morphology, with bright inner lobes
at a nuclear distance $\sim 2.5$~kpc ($\simeq 30\arcsec$), intermediate ridges
at a nuclear distance $\sim 15$~kpc ($\simeq 3\arcmin$) and an outer diffuse
``halo'' at a nuclear distance $\sim 40$~kpc ($\simeq 8\arcmin$) (we shall
refer to these as the ``inner'' lobes, ``intermediate'' ridges, and the outer 
halo, respectively). For reference, each ACIS chip is approximately $8\arcmin
\times 8\arcmin$, so we shall concentrate on the morphology on scales
comparable to the inner and intermediate radio structures. Nevertheless, the S3
and S2 chips combined cover almost all of the region corresponding to the outer
halo radio emission.

\begin{figure}

\vspace{-1.6cm}
\centerline{
\includegraphics[scale=0.31]{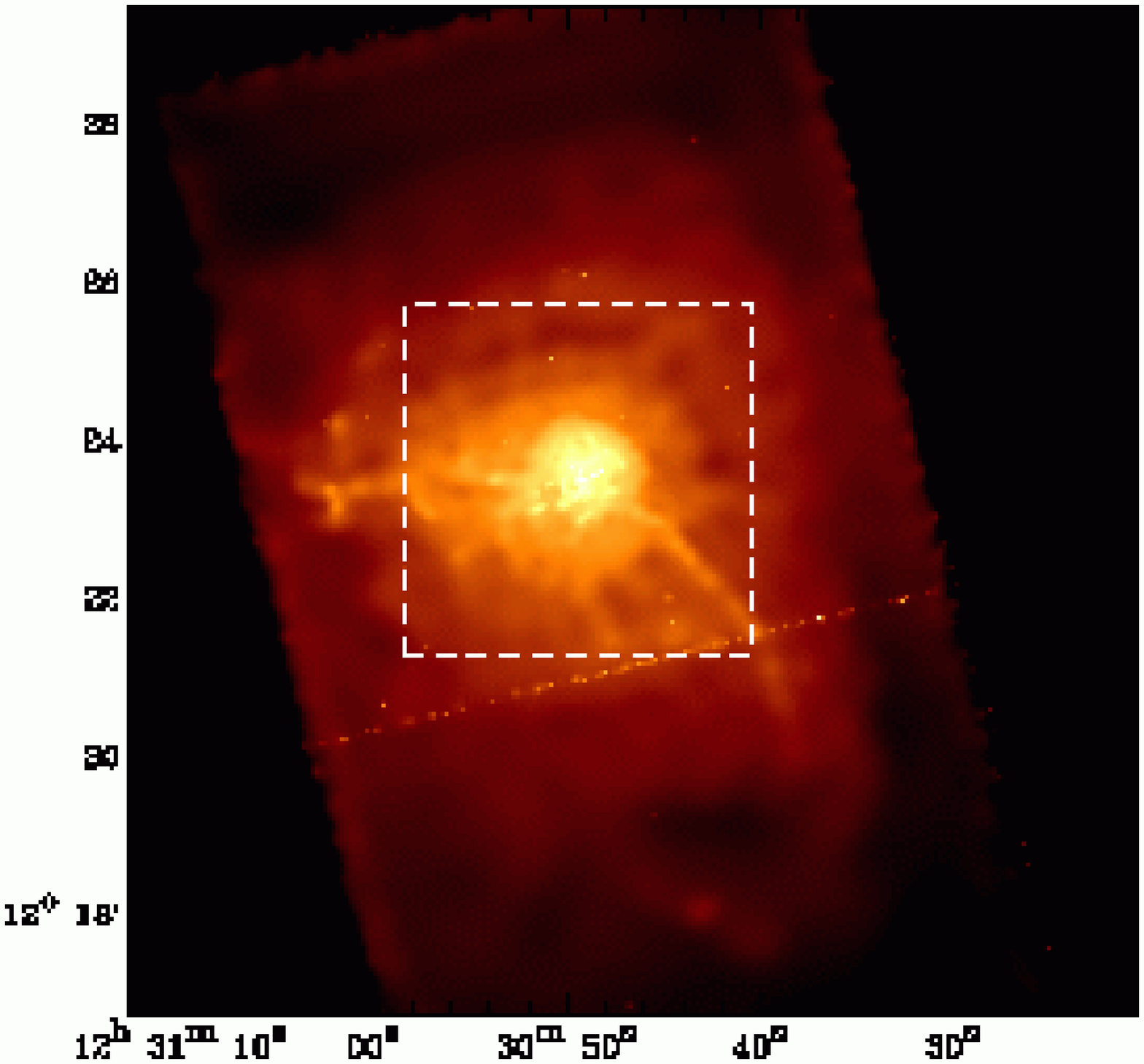}
\includegraphics[scale=0.31]{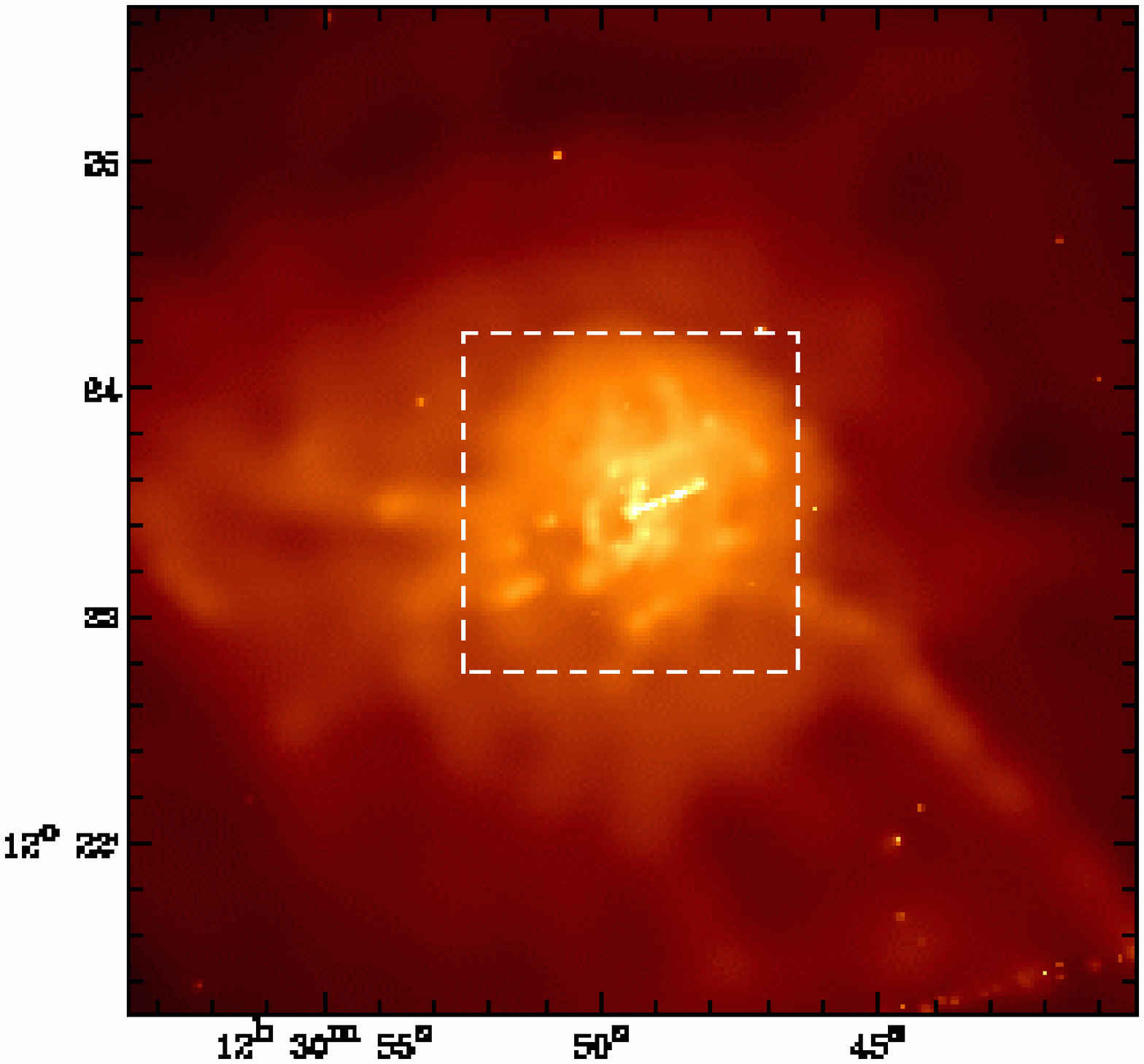}
}
\centerline{
\includegraphics[scale=0.675]{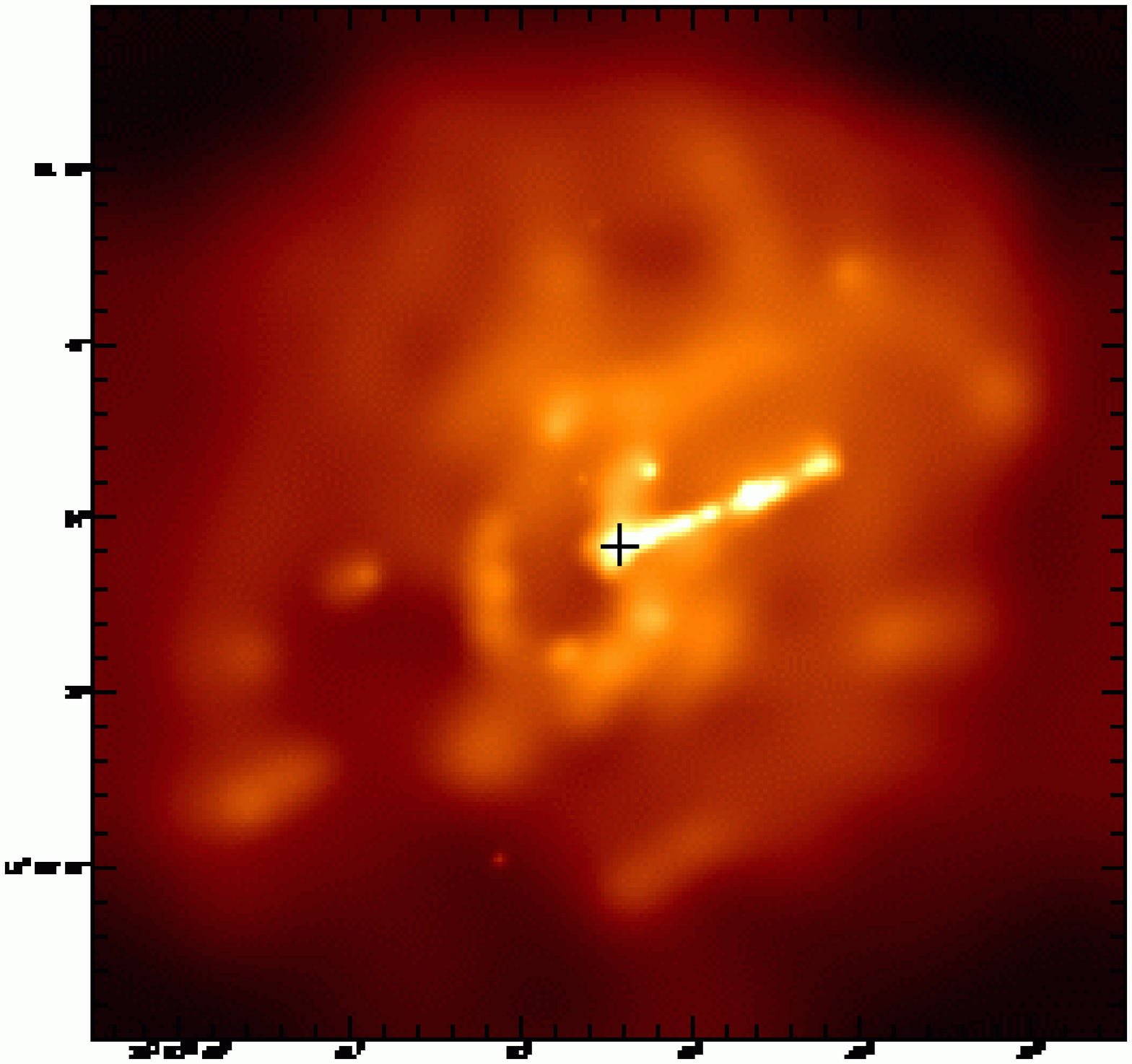}
}

\caption{\label{fig:cxo} A color representation of the \emph{Chandra} image of
  M87 and the core of the Virgo cluster in the 0.5~--~7~keV energy band. The
  images have been background subtracted, exposure map corrected and adaptively
  smoothed. The upper left panel shows the large scale structure and the upper
  right and bottom panels show enlargements of the regions outlined by dashed
  white lines in the upper left and upper right panels, respectively. The `+'
  sign indicates the nucleus of M87. A description of the morphology is given
  in Section~\ref{sec:morphology}.}

\end{figure}

Adaptively smoothed 0.5~--~7~keV X-ray images of the core of the Virgo cluster
are shown in Fig. \ref{fig:cxo} for three different spatial scales. There is
considerable structure on all scales from arcsecond to tens of arcminutes.
Concentrating on the central arcminute or so (bottom panel of Fig.
\ref{fig:cxo}), the brightest structure is the nucleus plus famous jet
extending towards the northwest (Wilson \& Yang 2002; Marshall et al. 2002).
The X-ray surface brightness in the central arcminute is very clumpy and a
number of point sources are visible. There are also major depressions in the
X-ray emission; one to the west of the nucleus just beyond the jet, one
immediately to the southeast of the nucleus, and another beyond that (i.e. 15
-- $30\arcsec$ from the nucleus) to the southeast and south.

\begin{figure}

\centerline{\includegraphics[scale=1.0,angle=0]{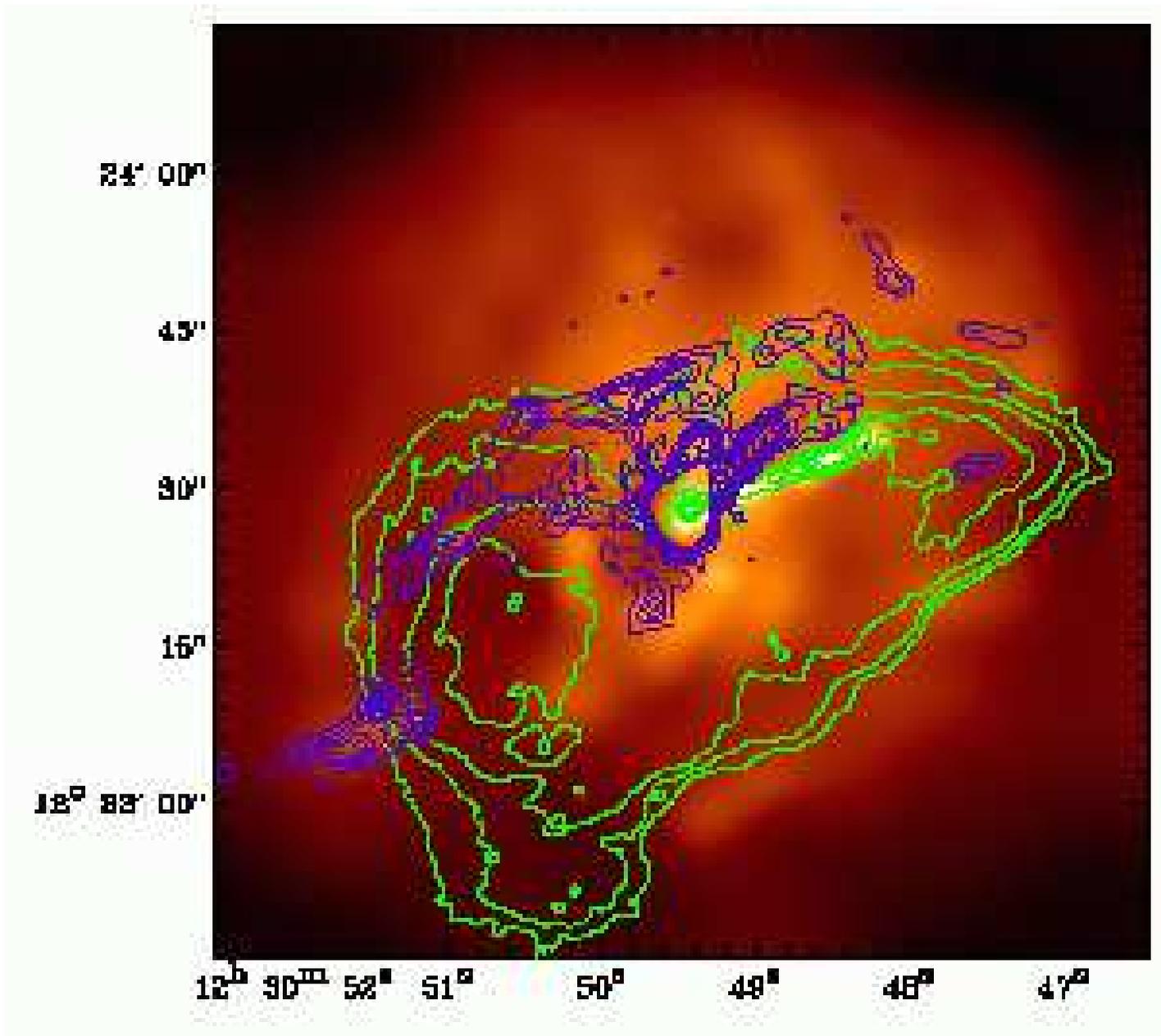}}

\caption{ \label{fig:cxo_radio_halpha} A color representation of the inner
  parts of M87 in the 0.5~--~7~keV energy band with contours of 6~cm radio
  emission (green) and H$\alpha$ + [N II] (blue) overlayed. The radio map has
  been smoothed by a Gaussian of FWHM $1\arcsec$, and the contours are
  logarithmically spaced. The H$\alpha$ + [N II] image has been smoothed by a
  Gaussian of FWHM $1\arcsec$, and the contours are linearly spaced. Note the
  similar nuclear distances of the inner radio lobes and the approximately
  circular ``front'' in X-rays (nuclear distance 45\arcsec).}

\end{figure}

A morphological comparison between the X-ray, 6~cm radio (Hines, Eilek \& Owen
1989) and H$\alpha$ + [N II] $\lambda\lambda 6548$, 6584 (Sparks, Ford \&
Kinney 1993) emission of the central $\simeq 1\farcm5$ is shown in Fig.
\ref{fig:cxo_radio_halpha}. The radio and X-ray emissions from the jet are well
correlated (Wilson \& Yang 2002; Marshall et al. 2002). The two inner radio
lobes, however, correspond to depressions in the X-ray surface brightness. Such
a phenomenon has been seen in other clusters, such as Hydra A (McNamara et al.
2000) and Perseus (Fabian et al. 2000), although on much larger spatial scales.
The H$\alpha$ + [N II] emission line gas is also excluded from the inner radio
lobes with the optical line-emitting gas predominantly on the north side of the
radio structure (Sparks 1999). Some knots in the emission line and X-ray gas
are associated with each other,  especially on the northern and eastern
boundaries of the inner eastern radio lobe.

As can be seen in Fig. \ref{fig:cxo} the decline in X-ray surface brightness
with increasing nuclear distance is not smooth. Instead, there are quite abrupt
``edges'' in brightness at nuclear distances of $\simeq$ 45\arcsec\ and
$\simeq$ 3\arcmin. These ``edges'' may also be seen in a plot of the
azimuthally averaged surface brightness vs. nuclear distance shown in Fig.
\ref{fig:beta}. It is notable that these two roughly circular features are just
outside the inner radio lobe and at comparable nuclear distance to the
intermediate radio ridges, respectively. This correspondence suggests the
``edges'' may have resulted from time variability in the active nucleus,
perhaps being associated with past epochs of jet production. In the model of
RHB, a short episode of jet activity eventually produces buoyant bubbles of
radio plasma that rise at $\sim 0.6 - 0.7$ times the sound speed, along with a
spherical ``pulse'' that expands at approximately the sound speed. We speculate
that the two roughly circular ``edges'' may correspond to the spherical pulses
associated with two earlier jet ejecta. While RHB discuss the results of their
simulation in the context of powerful FR-II radio sources, both types of
feature -- the buoyant bubbles and the spherical pulse -- are expected to be
present in the late-time evolution of a lower power (FR-I) radio source, such
as M87. This model is discussed further in Section~\ref{sec:fronts}.

The most striking feature on large scales is an X-ray ``arc'' running from
$\simeq$ 3\farcm3 east of the nucleus, across the central region of M87, to
$\simeq$ 4$\farcm9$ to the southwest (Figs \ref{fig:cxo} and
\ref{fig:color_cxo_radio}). On the east side of the nucleus, the arc
corresponds to a plume of radio emission (cf. Owen, Eilek \& Kassim 2000), and
ends at the center of the ``ear''-shaped eastern lobe of radio emission. On the
west side, the narrow arc passes to the south of the intermediate scale radio
ridges 2\arcmin\ -- 3\arcmin\ west of the nucleus. Further out, the arc follows
diffuse filaments of radio emission that connect the intermediate scale radio
ridges with the fainter, outer radio halo.

\section{The Intra-Cluster Medium} \label{sec:t_p_ct}

To get an idea of the spatial variation of the properties of the intra-cluster
medium we first constructed an emission-weighted temperature map. To do this,
images were made in soft (0.5 -- 0.9~keV), medium (0.9 -- 1.2~keV) and hard
(1.2 -- 4~keV) energy bands. The background subtracted data were then
adaptively binned (Sanders \& Fabian 2001) to ensure comparable fractional 
errors per bin in each band. Theoretical color ratio tables were computed as a
function of temperature and metalicity using the {\sc mekal} thermal plasma
model in {\sc xspec}, assuming absorption by only the Galactic column. The
color ratios of the data were compared with these theoretically computed color
ratios to construct a temperature map of the Virgo cluster (Fig.
\ref{fig:temp}).

A few things are immediately apparent from the emission-weighted temperature
map: (i) gas in the X-ray arc is cooler than surrounding intra-cluster gas at a
similar nuclear distance, (ii) there is a region of cooler gas immediately
surrounding the nucleus extending $\simeq 50\arcsec$ to the north and $\simeq
25\arcsec$ to the south, corresponding to the region of higher surface
brightness around the nucleus (see Fig.~\ref{fig:cxo}), and (iii) there is no
evidence of strong shock heating associated with the inner radio lobes.

Since a deprojection analysis was not performed, the temperature map shows the
emission-weighted mean temperature in each bin. If the gas distribution is
approximately spherically symmetric, with the density rapidly increasing
inwards, the properties of the gas in a given bin are dominated by the
properties of the gas at the same distance from the nucleus as the projected
distance of the bin from the nucleus. The azimuthally averaged brightness
profile of the inner part of the Virgo cluster is well described (i.e. to
within $10\%$ at all nuclear distances; see Fig. \ref{fig:beta}) by an
isothermal ``$\beta$-model'' (Sarazin \& Bahcall 1977) in which surface
brightness $S(r) \propto (1 + [r / r_c ]^2)^{-3\beta + 0.5}$ with a core radius
$r_c = 18\arcsec$ and $\beta = 0.4$, in good agreement with \emph{ROSAT} HRI
results (B\"ohringer 1999). Thus, the X-ray emission from the Virgo cluster is
strongly peaked towards the center with the possible exception of the inner
$\sim 20\arcsec$ and the X-ray arcs, so the inferred temperatures are a
reasonable approximation to the temperatures of gas at that physical distance
from the nucleus.

Motivated by the temperature map (Fig. \ref{fig:temp}), spectra were extracted
from a number of $25\arcsec \times 25\arcsec$ square regions, each containing
$\sim {\rm few} \times 10^3 - 10^4$ counts, to compare, in detail, the gas
properties in the arc and the ambient intra-cluster medium. An
emission-weighted instrument response was made for each region, and the
individual spectra were modeled by either (i) a single temperature VMEKAL
thermal plasma model or (ii) a two temperature VMEKAL model if the additional
component improved the $\chi^2$/d.o.f. by $> 0.04$. The abundances of O, Mg,
Si, S and Fe were allowed to vary but, if a two temperature model was used,
both components were constrained to have the same abundances.

The variation of temperature(s) with nuclear distance for different azimuthal
angles is shown in Fig.~\ref{fig:temp_vs_r}. The arc to the east and southwest
is much better described by a two temperature model, with temperatures of
$\approxgt 2$~keV and $\simeq 1$~keV. In contrast, the ambient gas is, in
general, better described by a  single temperature model with the temperature
rising outwards from $\simeq 1.5$~keV at 6~kpc to $\simeq 2.5$~keV at $\ge
20$~kpc. The region within $\simeq 6$~--9~kpc, corresponding approximately to
the apparently cooler gas immediately surrounding the nucleus (see
Fig.~\ref{fig:temp}), is also better described by a two temperature model, with
temperatures of $\simeq 1$~keV and $\approxgt 3$~keV. The properties of the gas
in the arc are clearly different from those in the surrounding cluster gas.

Focusing on the cooler region immediately surrounding the nucleus, we see that
the coolest gas in the central arcminute is well correlated with optical
H$\alpha$ + [N II] $\lambda\lambda 6548$, 6584 line emission, which is also
concentrated to the north of the nucleus (Sparks et al. 1993; our Fig.
\ref{fig:cxo_radio_halpha}). Moving farther out, there is another small
filament of H$\alpha$ + [N II] emitting gas located near the northwestern
corner of the ``ear'' (Gavazzi et al. 2000). This filament corresponds to a
region of increased X-ray surface brightness (Fig. \ref{fig:cxo}, upper left)
and cooler (Fig. \ref{fig:temp}) gas. The coincidence of the cooler X-ray gas
with the H$\alpha$ emitting gas suggests these may be regions in which mass is
dropping out of the hot intra-cluster medium.

To study the pressure and cooling time in a given bin, the emission measure was
converted to an electron density by assuming the path length through the bin is
equal to the distance of that bin from the nucleus, an assumption supported by
the rapid increase of X-ray surface brightness towards smaller nuclear
distances (see above discussion). The electron density (and gas pressure) are
$\propto 1 / \sqrt{l}$, where $l$ is the path length through the gas so a
factor of 2 error in the path length would correspond to only a factor of
$\sqrt{2}$ error in these quantities. To minimize the error in the assumed path
length, we treated the arcs to east and southwest separately and assumed the
path length through an arc is 1 kpc for those regions of the arc at a nuclear
distance $< 4\arcmin$ ($< 18$ kpc). The cooling time was computed using the
cooling function of Sarazin \& White (1987) for solar metallicity gas,
correcting for the printing error noted by Westbury \& Henriksen (1992). The
rate of cooling ($\Lambda(T)$) of solar metallicity gas is faster than zero
metallicity gas by factors of $\simeq 1.5$, 2 and 6 at temperatures of 10, 2
and 1 keV, respectively.

The electron density, total gas pressure and cooling time are shown in
Fig.~\ref{fig:d_p_ct} as a function of nuclear distance for a number of
different azimuthal angles. Away from the arc, the density and pressure decline
rapidly with increasing nuclear distance while the cooling time rises. The
properties of the gas away from the arc are similar to those previously derived
for the cluster as a whole, with the run of temperature with nuclear distance
being consistent with \emph{XMM-Newton} (B\"ohringer et al. 2001) and the run
of density with nuclear distance being consistent with \emph{ROSAT} (Nulsen \&
B\"ohringer 1995). The gas in the arcs, however, is denser and probably
over-pressured with respect to the surrounding cluster gas at a similar nuclear
distance; a deprojection analysis of the ambient cluster emission would provide
a more accurate estimate of the pressure of the ambient gas, but would be
limited by the lack of symmetry of the X-ray emission. In addition, the density
and pressure decline more slowly with increasing nuclear distance in the arc
than in the surrounding cluster gas, although the slope of this decline is
sensitive to the assumed path length through the gas in the arc.

\section{Metal Abundances} \label{sec:abund}

A knowledge of the metal abundances of the cooler gas in the arc and the hotter
surrounding cluster gas is important for differentiating between various
theoretical models of the arc (see Section~\ref{sec:arc}). To study the metal
abundances to the east of the nucleus, counts were extracted from two
$25\arcsec \times 25\arcsec$ square regions, one centered on the cooler gas
$1\farcm6$ east of the nucleus in the arc (which we refer to as the ``arc''
gas), and another over the nearby cluster gas $1\farcm6$ north of that (which
we refer to as the ``ambient'' gas). There are 8667 and 3323 cts in these two
regions, respectively. In order to model the spectra, an emission-weighted mean
instrument response was used. This procedure does not provide a perfect
description of the instrument response, with residuals corresponding to the
iridium M-edge (1.8~--~2.2~keV, as noted by Markevitch and Vikhlinin [2001]),
and at energies corresponding to uncertainties in ACIS gain maps (1.6 and
2.6~keV\footnote{ACIS calibration issue dated 3 Jan 2002 at
\url{http://asc.harvard.edu/cal/Links/Acis/acis/Cal\_projects/index.html}}),
the effect of which is to increase both the $\chi^2$ values of our fits and the
uncertainty in our determination of the Si and S abundances that have strong
emission lines in these energy ranges. The ambient gas to the northeast of the
nucleus may be described by a single temperature thermal plasma model (the
VMEKAL model in XSPEC) in which the abundances of O, Mg, Si, S and Fe are
allowed to vary, with $\chi^2$/d.o.f. = 134/110. The parameters of this model
are given in Table~\ref{tab:spec} and the data and model are shown in the left
panel of Fig.~\ref{fig:spec}. The spectrum of the arc to the east of the
nucleus is more complicated, and is poorly described by a single temperature
thermal plasma model, which has a $\chi^2$/d.o.f. = 254/136. The arc can be
described by a two temperature thermal plasma model, with temperatures of
1.77~keV and 1.07~keV. There is a degeneracy between the relative
normalizations and metalicities of the two components, and this can be broken
by assuming the warmer component has the same parameters as the ambient gas.
The cooler component is then found to have enhanced Mg, Si, S and Fe abundances
relative to the ambient gas and comparable O abundance. The model parameters
are given in Table~\ref{tab:spec} and the model fit shown in the right panel of
Fig.~\ref{fig:spec} with a $\chi^2$/d.o.f. = 188/136.

To the west of the nucleus, counts were extracted from six $25\arcsec \times
25\arcsec$ square regions, three centered on the cooler gas in the arc
$1\farcm6$ southwest of the nucleus, and three over the nearby cluster gas a
similar distance west of the nucleus. There are 18065 and 10815 cts in the
three regions over the arc and the three regions over the ambient gas,
respectively. These extraction regions are larger than those used to the east
of the nucleus because we found that a greater S/N ratio was required to
discern variations in the metal abundances. The spectra of the arc and the
nearby cluster gas were then analyzed in the same way as the spectra from the
regions east of the nucleus. The ambient gas to the west of the nucleus can be
described by a single temperature thermal plasma model (see
Table~\ref{tab:spec}) with $\chi^2$/d.o.f. = 440/362. The spectrum of the arc
to the southwest of the nucleus is not well described by a single temperature
plasma model ($\chi^2$/d.o.f. = 683/396). The arc can be described by a two
temperature thermal plasma model, with temperatures of 2.09~keV and 1.08~keV.
If the warmer component has the same parameters as the ambient gas, the cooler
component is found to have enhanced Mg and Si abundances than the ambient gas,
comparable O and Fe, and comparable or lower S abundance. The model parameters
are given in Table~\ref{tab:spec} with a $\chi^2$/d.o.f. = 557/395.

To summarize, the physical properties of the gas in the X-ray arc are
significantly different to those of the ambient cluster gas, in particular an
additional cooler thermal component is present. The metal abundances of some of
the heavy elements in the cooler gas in the arc, both to the east and west, are
consistent with being enhanced relative to those of the ambient gas. Our
conclusions are in broad agreement with the \emph{XMM-Newton} observations of
M87 (Belsole et al. 2001).

\section{Discussion} \label{sec:discussion}

\subsection{An Estimate of the Minimum Power of the Jet}

If we assume the ``holes'' in the X-ray gas that correspond to the inner radio
lobes (see Fig. \ref{fig:cxo_radio_halpha}) are cavities created by the radio
jet then we can make a crude estimate of the minimum power of the jet. The
cavities are approximately spherical with typical diameters $\simeq$ 1.5 kpc
and volumes  $V = 6 \times 10^{64}$~cm$^{3}$, and the pressure in the regions
around the cavities is $\sim$ 2 $\times$ 10$^{-9}$ dyn cm$^{-2}$. An estimate
of the age of a cavity can be obtained from the free-fall timescale, $t_{\rm
ff} = R / c_{\rm s}$ where $R$ is the radius of the cavity, $R = 2.3 \times
10^{21}$~cm, and $c_{\rm s}$ is the sound speed, $c_{\rm s} = 3.4 \times 10^7
(T / 1 {\rm ~keV})^{1/2}$~cm~s$^{-1}$. Assuming the gas temperature $T = 1$~keV
gives a jet power $F_{\rm jet} \simeq 2PV / t_{\rm ff} = 3 \times
10^{42}$~erg~s$^{-1}$. If the cavities are evacuated faster than $c_{\rm s}$,
then $F_{\rm jet}$ would be higher, although there appear to be no signs of
sharp edges in the X-ray image that might be shocks. Our estimate of $F_{\rm
jet}$ is at the lower end of previous estimates (e.g. Bicknell \& Begelman
[1999] find $F_{\rm jet} \simeq 10^{43-44}$~erg~s$^{-1}$), although there are
inevitably uncertainties in our estimates of $t_{\rm ff}$ and other quantities.
The cavity cannot be inflated much more slowly than the above estimate or it
would have risen too far from the nucleus through buoyancy effects. The bubbles
are expected to rise at $\sim 0.6 - 0.7 c_{\rm s}$ (Churazov et al. 2001), and
cannot have traveled more than 2~kpc giving a maximum age of $\sim 4 \times
t_{\rm ff}$ and $F_{\rm jet} \approxgt 10^{42}$~erg~s$^{-1}$.

\subsection{The X-ray ``Fronts''} \label{sec:fronts}

\begin{figure}

\centerline{\includegraphics[scale=0.9,angle=270]{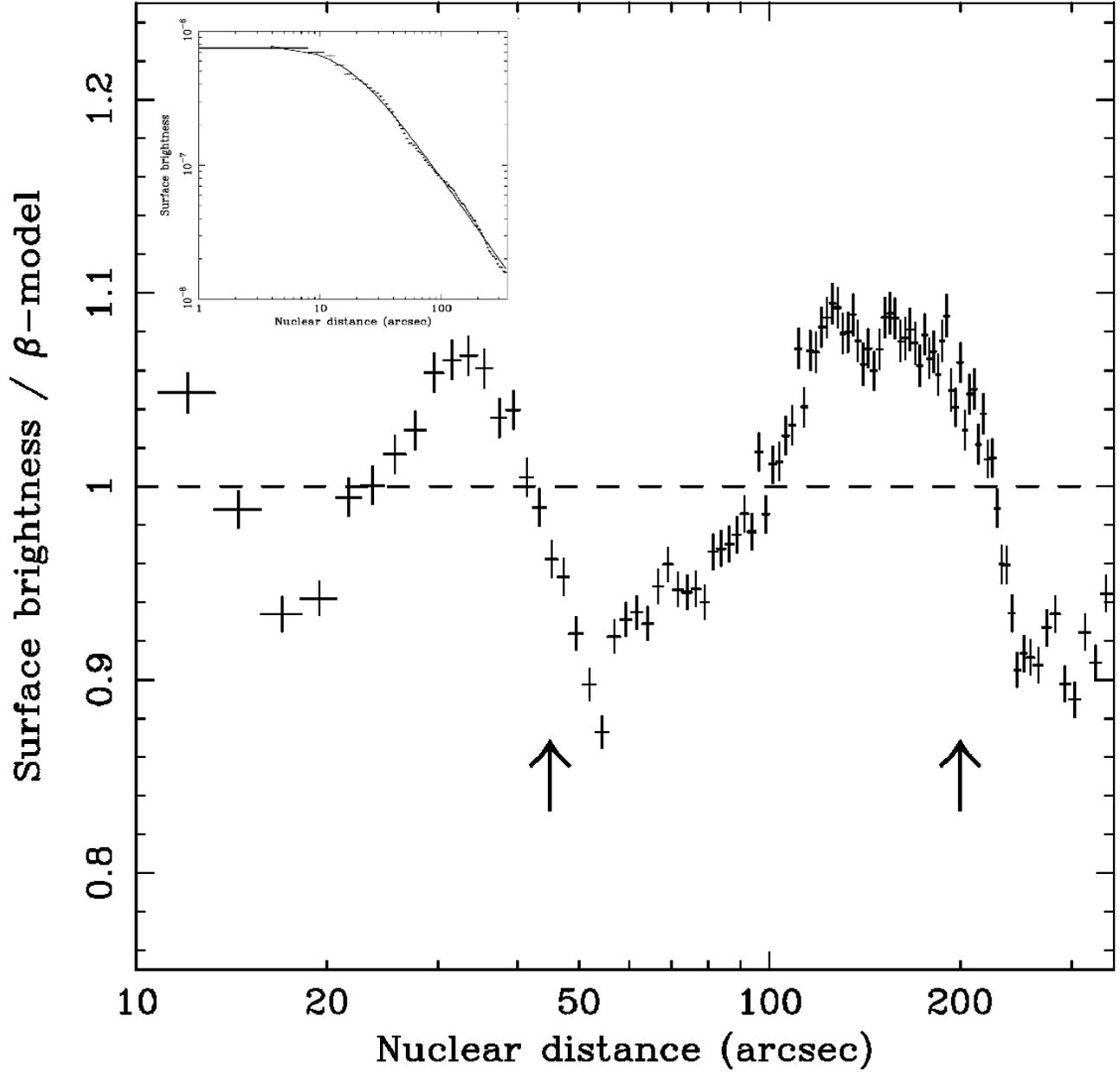}}

\caption{\label{fig:beta} A plot showing the ratio of the azimuthally averaged
  X-ray surface brightness to the best fitting ``$\beta$-model'' (see
  Section~\ref{sec:t_p_ct}) as a function of nuclear distance. There are
  systematic deviations of $\pm 10\%$ in surface brightness from the
  $\beta$-model fit (shown as the inset). Arrows indicate the approximate
  locations of the sudden  declines (``edges'' or ``fronts'') in surface
  brightness discussed in the text. The X-ray emission from the  nucleus and
  jet were excluded from this analysis. }

\end{figure}

\begin{figure}

\centerline{\includegraphics[scale=1.0,angle=0]{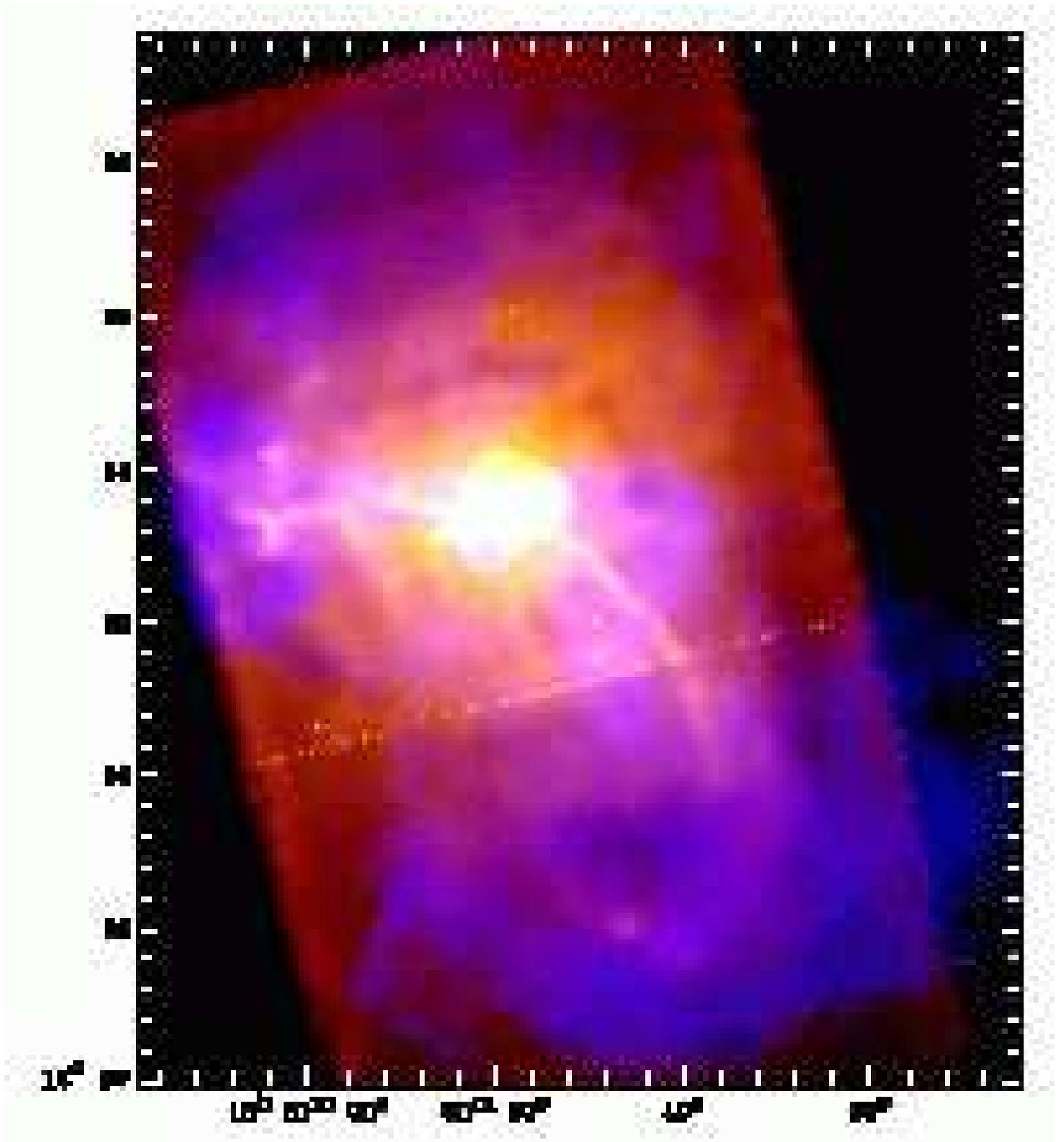}}

\caption{ \label{fig:color_cxo_radio} An X-ray -- radio color overlay of the
  Virgo cluster. The adaptively smoothed \emph{Chandra} 0.5 -- 7~keV X-ray
  image is red/yellow, while the 90~cm radio emission (made from VLA archival
  data) is blue. Note the similar nuclear distances of the intermediate radio
  ridges and the approximately circular ``front'' in X-rays (nuclear distance
  $\simeq 3\arcmin$)}

\end{figure}

\begin{figure}

\centerline{\includegraphics[scale=0.8,angle=0]{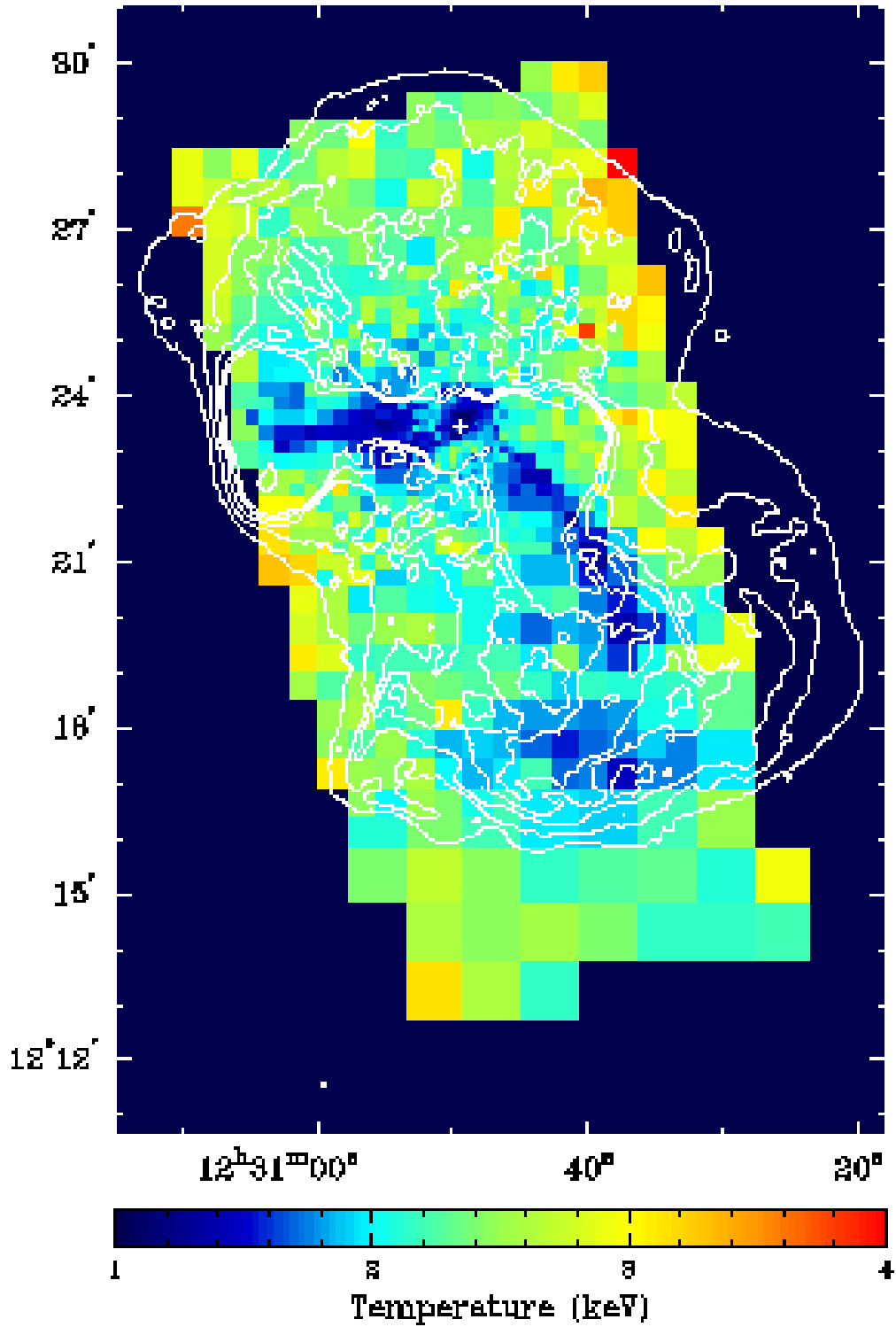}}

\caption{ \label{fig:temp} Temperature map of the core of the Virgo cluster.
  This map was constructed from adaptively binned \emph{Chandra} images, using
  the technique described in Section ~\ref{sec:t_p_ct}. Overlayed are contours
  of 90~cm radio emission (made from VLA archival data). The contours are at
  0.01, 0.05, 0.1, 0.15 and 0.2 Jy beam$^{-1}$.}

\end{figure}

\begin{figure}
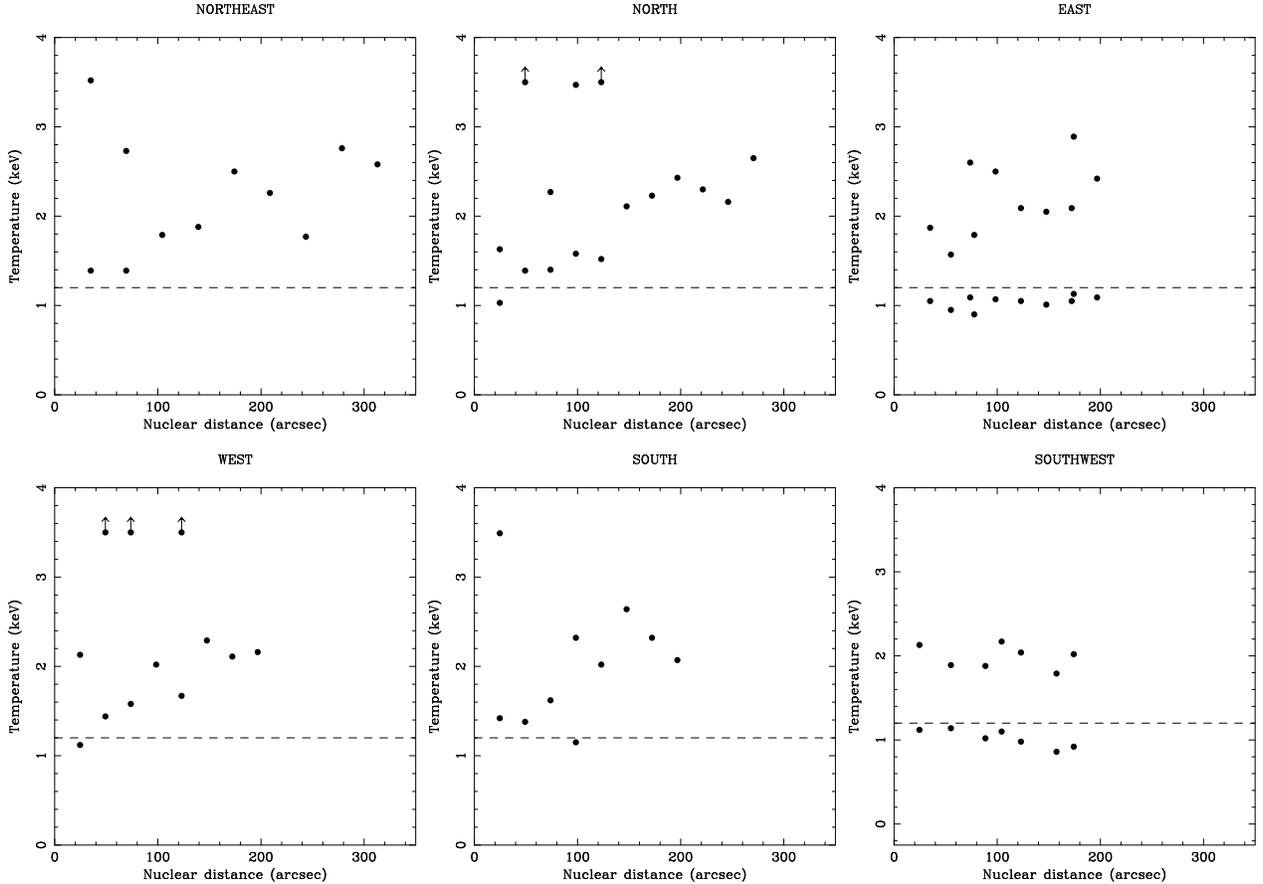


\centerline{
  \includegraphics[angle=270,scale=0.3]{fig6a.ps}
  \includegraphics[angle=270,scale=0.3]{fig6b.ps}
  \includegraphics[angle=270,scale=0.3]{fig6c.ps}
}

\vspace{0.25cm}

\centerline{
  \includegraphics[angle=270,scale=0.3]{fig6d.ps}
  \includegraphics[angle=270,scale=0.3]{fig6e.ps}
  \includegraphics[angle=270,scale=0.3]{fig6f.ps}
}

\caption{\label{fig:temp_vs_r} Plots of temperature vs. nuclear distance for
  different azimuthal angles corresponding to the ambient cluster medium
  (towards the northeast [top left], north [top center], west [bottom left] and
  south [bottom center]) and the arc (towards the east [top right] and
  southwest [bottom right]). At a given nuclear distance and azimuthal angle
  the spectrum is described by a single temperature thermal plasma model unless
  a two temperature thermal plasma model results in an improvement in
  $\chi^2$/d.o.f. of $> 0.04$, in which case a two temperature thermal plasma
  model is used.}

\end{figure}

\begin{figure}
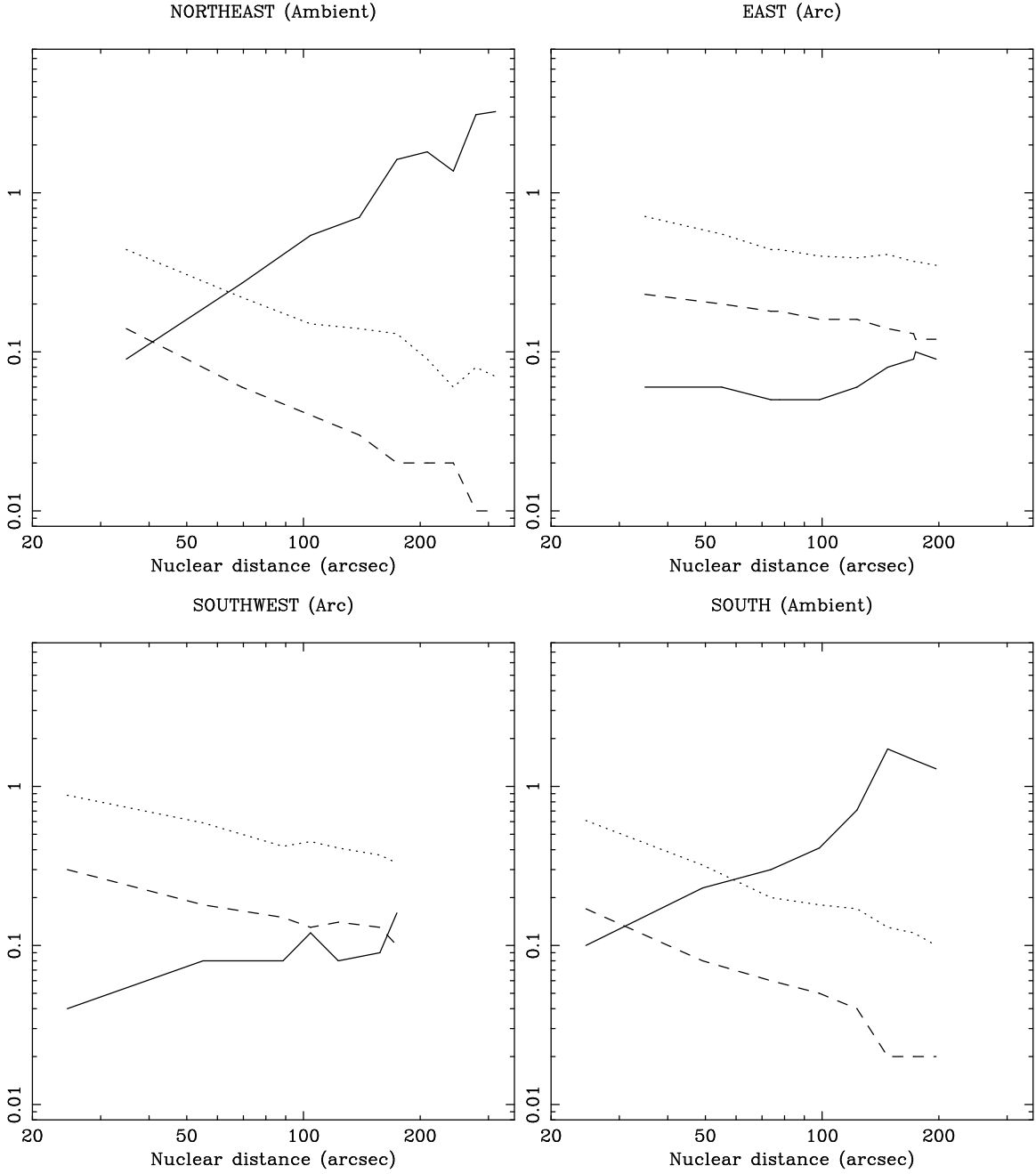


\centerline{
  \includegraphics[angle=270,scale=0.45]{fig7a.ps}
  \includegraphics[angle=270,scale=0.45]{fig7b.ps}
}

\vspace{0.25cm}

\centerline{
  \includegraphics[angle=270,scale=0.45]{fig7c.ps}
  \includegraphics[angle=270,scale=0.45]{fig7d.ps}
}

\caption{\label{fig:d_p_ct} Plots of electron density (cm$^{-3}$; dashed line),
  total gas pressure (cm$^{-3}$ keV; dotted line) and cooling time (Gyr; solid
  line) vs. nuclear distance for different azimuthal angles corresponding to
  the ambient cluster medium (towards the northeast [top left] and south
  [bottom right]) and the arc (towards the east [top right] and southwest
  [bottom left]).}

\end{figure}

\begin{figure}
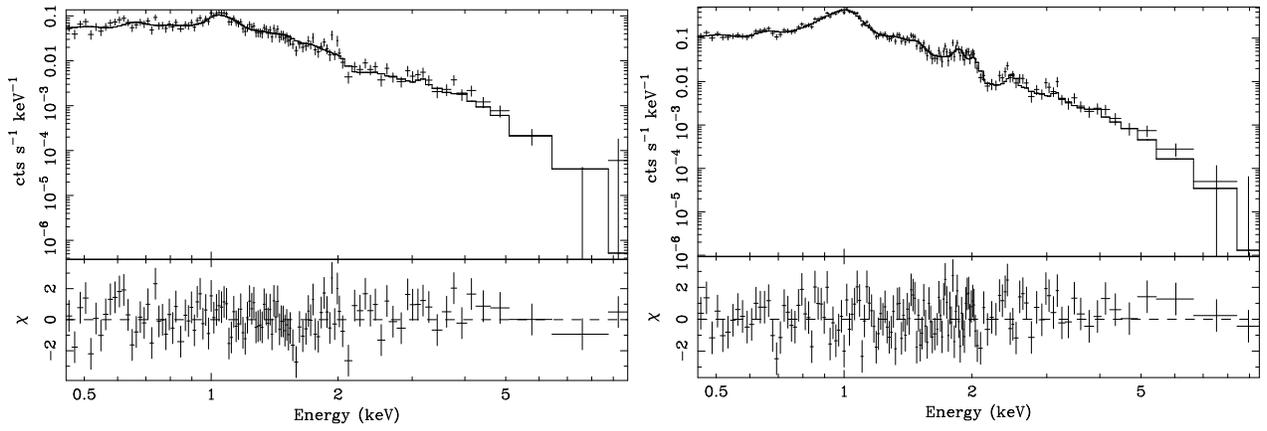


\centerline{
  \includegraphics[scale=0.35,angle=270]{fig8a.ps}
  \includegraphics[scale=0.35,angle=270]{fig8b.ps}
}

\caption{ \label{fig:spec} Spectra of the ambient cluster gas to the northeast
  of the nucleus (left panel) and in the arc to the east of the nucleus (right
  panel). Both spectra were extracted from $25\arcsec \times 25\arcsec$ square
  regions. The spectrum of the gas in the arc may be approximated by the sum of
  the spectrum of the ambient gas and a cooler, more metal rich component, and
  this sum is plotted in the right hand panel. See Section~\ref{sec:abund} for
  more details.}

\end{figure}

The structure in the radial distribution of X-ray surface brightness (see Figs
\ref{fig:cxo}~--~\ref{fig:color_cxo_radio}) could be some effect inherent to a
cooling flow. However, we have noted (Section \ref{sec:morphology}) that the
two most noticeable sharp declines in azimuthally averaged X-ray surface
brightness are just outside the inner radio lobes and at comparable nuclear
distance to the intermediate radio ridges. This similarity between the nuclear
distances of the radio lobes and the declines in surface brightness suggests
instead that the ``edges'' may be ``fronts'' in the cluster atmosphere created
by earlier ejecta from the active nucleus of M87.

We consider a scenario based on the results of numerical hydrodynamical
simulations by RHB of the effects of the jets on the intra-cluster medium. RHB
discuss their simulation results in the context of powerful FR-II radio
galaxies but the end results are expected to be similar in a weaker FR-I radio
galaxy. The end effect of the transient jet activity is that the cluster
expands in response to heating of the cluster core. An approximately spherical
pulse propagates outwards at the sound speed (see Fig.~6 of RHB), ultimately
leaving the core of the cluster at a fractionally lower density and a higher
temperature. The amplitude of the density wave in the RHB simulation is $\pm
10\%$ which is dependent upon the initial conditions but is large enough to
produce the $\simeq 10\%$ fluctuations seen in the azimuthally averaged X-ray
surface brightness when compared to the best fitting ``$\beta$-model'' shown in
Fig. \ref{fig:beta}. The fact that the two fronts are at slightly larger
nuclear distances than the inner radio lobes and the intermediate ridges
suggests that each of these radio features is in the ``passive'' or buoyant
bubble phase that characterizes the late stage of evolution. Such would be
consistent with the absence of strong shocks around the cavities  inflated by
the inner radio lobes. The $\simeq 2\farcm5$ (12 kpc) separation of successive
fronts corresponds to a sound crossing time of $t_{\rm sc} = 2 \times 10^7 (T /
2.5 {\rm ~keV} )^{-1/2}$~yrs, and indicates there may be relatively brief
periods of jet activity every $10^7$~yrs or so. If this is the case, the
time-averaged energy injected into the cluster by the jet may be significantly
lower than the present instantaneous power of the jet. Further detailed
numerical simulations, tailored to the environment of M87, are required to
determine whether such a model can indeed reproduce the ``fronts'' and what
conditions are needed. In particular, such simulations are required to find out
what effect the putative pulse has on the cluster core. The model of RHB also
predicts subtle (tens of per cent) changes in the gas temperature associated
with the changes in X-ray surface brightness, and while these changes are too
small to be detected in our data set, one could look for them in a future
observation with a significantly longer integration time.

\subsection{The X-ray Arcs} \label{sec:arc}

Our observations of the X-ray arcs to the east and southwest and associated
with the radio ridges show that they are narrow, with a FWHM of only $\simeq
14\arcsec$ (1 kpc). The length to width ratio of the southwestern arc is
$\simeq 20$. We have also found (Section \ref{sec:t_p_ct}) that the gas in the
arcs is cooler than the surrounding cluster gas, in agreement with previous
studies, and appears to contain two components with different temperatures. The
cooler component may have some metal abundances enhanced with respect to the
warmer gas in the cluster (Table~\ref{tab:spec}, Section~\ref{sec:abund}).

A number of models have been proposed to explain these arcs. Firstly, it should
be noted that although the X-ray emissivity of the gas may be enhanced by
compression, shocks or cosmic ray heating processes tend to raise the gas
temperature rather than lower it, and can be ruled out (B\"ohringer et al.
1995). Another possibility is that, as the radio ejecta (which are likely to be
in the late ``buoyant plume'' phase [RHB]; cf. Section~\ref{sec:fronts}) pass 
through a multi-phase cooling flow, they disrupt the ``cold'' phase (consisting
of cold, denser clouds) and mix it with the ``hot'' phase. This causes a rapid
cooling of the hot phase and an enhancement of the density (B\"ohringer et al.
1995). However, whether the ambient intra-cluster medium is, in fact,
multi-phase remains controversial (we are able to adequately describe most of
the intra-cluster gas as a single temperature plasma). A further possibility is
that cooler, low entropy gas near the nucleus is mixed into the radio plasma,
generating buoyant bubbles of gas (Churazov et al. 2001; Br\"uggen et al. 2002;
RHB). These bubbles would rise in the cluster gravitational field and expand
until they have a  density equal to that of the surrounding medium.

The in-situ disruption of gas in a cold phase, followed by mixing, should leave
cooler gas with metal abundances comparable to the surrounding cluster gas.  On
the other hand, if the cool gas was entrained at a small nuclear distance and
carried outwards,  the buoyant bubbles of cold gas should have metal abundances
appropriate to the cluster core. As discussed above (Section~\ref{sec:abund}),
we find that the cooler gas (see Table~\ref{tab:spec}) in the arc probably has
a higher metalicity for at least some elements than the ambient gas. These
higher abundances favor the buoyant bubble scenario. The morphology of the
X-ray arcs (i.e. narrow and well collimated) is consistent with the Churazov et
al. (2001) and RHB models in which a narrow column of dense gas is created.
The  numerical simulations of both Churazov et al. (2001) and RHB assumed
axi-symmetry which may artificially stabilize narrow structures elongated along
the  symmetry axis, although reasonably narrow, dense structures with an aspect
ratio of $\sim 10$ are also produced in the fully 3D simulations of Br\"uggen
et al. (2002). The sound-crossing time from the nucleus to a nuclear distance
of $200\arcsec$ is a ${\rm few} \times 10^7$ yrs, which is shorter than the
cooling time of gas in the arcs, which is $\approxgt 5 \times 10^7$ yrs and
closer to $10^8$ yrs at a nuclear distance of $200\arcsec$ (see
Fig.~\ref{fig:d_p_ct}). This means that there is time for cool gas entrained by
a buoyant bubble in the nucleus to be carried to the end of the arc before it
has cooled completely. If this were not the case, i.e. if the cooling time of
gas in the arcs was shorter than the sound-crossing time, then some in-situ
heating mechanism would be required, or the cool gas must have been transported
faster than the sound speed. The simulation by RHB produces both elongated,
buoyant plumes which lift gas from the center of the cluster and approximately
spherical pulses traveling at the sound speed that we have identified with the
X-ray ``fronts''. In this sense, the structure of the intra-cluster medium in
the inner part of the Virgo cluster conforms qualitatively to the expected
effects of repeated  injections of energy over a period of 10$^{8}$ yrs by
radio jets from the nucleus of Virgo A.

\section{Conclusions}

We have presented the highest resolution X-ray image to date of the core of the
Virgo cluster. Structure is observed on all spatial scales, from arcsecond
($\sim 100$ pc) to ten arcminutes ($\sim 50 $ kpc).

The inner radio lobes are aligned with depressions in the X-ray surface
brightness and there is no evidence of shock heating in the X-ray emission
immediately surrounding the inner radio lobes, suggesting that the radio plasma
has gently pushed aside the X-ray emitting gas. These cavities cannot have been
inflated much slower than the sound speed, however, or they would have risen
too far from the nucleus due to buoyancy effects. We estimate the jet power to
be $F_{\rm jet} \simeq 3 \times 10^{42}$ erg s$^{-1}$.

On larger scales the most striking feature is the X-ray arc running from the
east, across the central regions of M87, and off to the southwest. The gas in
the arc has at least two temperatures, with one component at the temperature of
the ambient intra-cluster medium and a cooler component at $\simeq 1$ keV. The
gas in the arcs is probably over-pressured with respect to, and somewhat more
metal rich than, the ambient intra-cluster medium.

Abrupt changes in surface brightness or ``fronts'' are seen at nuclear
distances slightly larger than the nuclear distances of the inner radio lobes
and intermediate radio ridges. Within the inner front, at nuclear distances
$\approxlt 45\arcsec$ ($\approxlt 3.5$ kpc) the gas has at least two
temperatures, with the cooler component at $\simeq 1$ keV, similar to the X-ray
arc. This cooler region is concentrated more to the north than the south of the
nucleus and is correlated with the H$\alpha$ + [NII] emission-line
distribution.

We suggest that a model based on the hydrodynamical simulations of RHB, scaled
to a lower power radio source such as M87, may explain the observed phenomena.
Intermittent jet activity has two effects on the cluster. Firstly, it inflates
buoyant bubbles of radio plasma that trail cold gas from the central regions in
their wakes as they rise at $\simeq 0.6$ -- 0.7 times the sound speed, thereby
producing the X-ray arcs. The gas dredged up from the nucleus is expected to
have higher metal abundances than the intra-cluster medium at large nuclear
distances. Secondly, at late times in the evolution of the radio source, the
injection of energy into the cluster core produces a ``pulse'' that expands at
the sound speed into the intra-cluster medium as the cluster adjusts to find a
new equilibrium. We identify this pulse with the observed X-ray ``fronts''.
Detailed numerical simulations tailored to the Virgo cluster are required to
explore this hypothesis.

\acknowledgments

We thank W. B. Sparks for providing the H$\alpha$ + [N II] map in electronic
form and NRAO for supplying the u-v data used to make the 90~cm map. We also
thank Chris Reynolds for discussions. This research was supported by NASA
grants NAG 81027 and NAG 81755. We are also grateful to D. Harris and S. Virani
of the \emph{Chandra} Science Center for their assistance with the
observations.



\begin{deluxetable}{ccccc}

\tablewidth{0pc} \rotate

\tablecaption{ X-ray Spectral Models of the ``Arc'' and Ambient Cluster Gas
  \label{tab:spec} }

\tablehead{

\colhead{Parameter} & \colhead{Northeast (ambient)} & \colhead{East (arc)} &
\colhead{West (ambient)} & \colhead{Southwest (arc)} \\

\colhead{} & \colhead{} & \colhead{= Northeast (ambient) +} & \colhead{} &
\colhead{= West (ambient) +}

}

\startdata

Counts & 3323 & 5344 & 10815 & 7250 \\

$N_{\rm H}$ ($\times 10^{20}$ cm$^{-2}$) & $8.8^{+2.2}_{-1.9}$ &
$3.6^{+3.4}_{-1.7}$ & $6.1^{+1.1}_{-1.1}$ & $1.1^{+4.3}_{-1.1}$ \\

$kT$ (keV) & $1.77^{+0.24}_{-0.11}$ & $1.07^{+0.02}_{-0.02}$ &
$2.09^{+0.08}_{-0.10}$ & $1.08^{+0.00}_{-0.02}$ \\

O & $0.55^{+0.40}_{-0.35}$ & $0.48^{+0.52}_{-0.37}$ & $0.59^{+0.27}_{-0.26}$ &
$0.76^{+0.62}_{-0.39}$ \\

Mg & $0.00^{+0.37}_{-0.00}$ & $1.28^{+1.28}_{-0.67}$ & $0.00^{+0.32}_{-0.00}$ &
$1.06^{+1.04}_{-0.62}$ \\

Si & $0.06^{+0.26}_{-0.06}$ & $2.13^{+1.29}_{-0.62}$ & $0.91^{+0.26}_{-0.24}$ &
$1.82^{+1.02}_{-0.71}$ \\

S & $0.19^{+0.39}_{-0.19}$ & $1.73^{+1.62}_{-0.72}$ & $1.46^{+0.40}_{-0.39}$ &
$0.37^{+1.09}_{-0.37}$ \\

Fe & $0.28^{+0.15}_{-0.08}$ & $1.08^{+0.39}_{-0.31}$ & $0.72^{+0.12}_{-0.10}$ &
$0.81^{+0.40}_{-0.29}$ \\

K\tablenotemark{a} & $4.4 \times 10^{-4}$ & $2.4 \times 10^{-4}$ & $11 \times
10^{-4}$ & $3.3 \times 10^{-4}$\\

$\chi^2$/d.o.f. & 134/110 & 188/136 & 440/362 & 557/395 \\

\enddata

\tablenotetext{a}{{\sc vmekal} normalization $K = 10^{-14} (4 \pi D_A^2 [1+z]^2
  )^{-1} \int n_e n_H dV$ where $D_A$ is the angular size distance (cm), $n_e$
  is the electron density (cm$^{-3}$) and $n_H$ is the hydrogen density
  (cm$^{-3}$).}

\end{deluxetable}

\vfil\eject


\end{document}